\documentclass[12pt]{article} 

\usepackage[a4paper,total={170mm,257mm},left=20mm,top=20mm]{geometry}
\usepackage{amsmath}
\usepackage{stackengine}
\usepackage{dirtytalk}

\usepackage{graphicx} 

\usepackage{xcolor}
\usepackage{setspace} 
\usepackage{array} 
\usepackage{paralist} 
\usepackage{verbatim} 
\usepackage{subfig} 
\usepackage{amsthm} 
\usepackage{amssymb}
\usepackage{fancyhdr} 
\usepackage{hyperref}

\usepackage[nottoc,notlof,notlot]{tocbibind} 
\usepackage[titles,subfigure]{tocloft} 

\usepackage{ marvosym }

\theoremstyle{definition}

\numberwithin{equation}{section}

\title{Virasoro Constraint for Uglov Matrix Model}
\author{\textsc{Osama Khlaif}$^{1,2}$ and \textsc{Taro Kimura}$^1$} 
\date{} 

\begin{document}

\maketitle

\vspace{-2.5em}
\begin{center}
$^1$Institut de Mathématiques de Bourgogne, Université Bourgogne Franche-Comté, France\\[.5em]
$^2$School of Mathematics, University of Birmingham, UK
\end{center}


\begin{abstract}
We study the root of unity limit of $(\textbf{q}, \textbf{t})$-deformed Virasoro matrix models, for which we call the resulting model Uglov matrix model.
We derive the associated Virasoro constraints on the partition function, and find agreement of the central charge with the expression obtained from the level-rank duality associated with the parafermion CFT.
\end{abstract}

\tableofcontents
\vspace{1em}
\hrule

\section{Introduction}

In recent years, matrix models have been playing a crucial role in studying gauge and string theories. Via the supersymmetric localization principle first applied in \cite{Pestun:2007rz}, partition functions of several supersymmetric gauge theories were reduced into matrix models that could be studied more easily. See a review article~\cite{Pestun:2016zxk} for more details. As a result of this reduction, many correspondences have been established between gauge theories and conformal field theories (CFTs) in 2d hence making the calculations of several physical observables, e.g., Wilson loops, more tractable, along the line of the BPS/CFT correspondence~\cite{Nekrasov:2015wsu}.

One interesting correspondence that has been studied lately is a 3d version of the BPS/CFT correspondence. It was found that the matrix model form of the 3d holomorphic blocks associated with $\mathcal{N}=2$ 3d supersymmetric Chern-Simons-Yang-Mills matter theory on $\mathbb{D}^2\times_{q} \mathbb{S}^1$ is identified as that associated with the $(\textbf{q},\textbf{t})$-deformed Virasoro algebra \cite{Pasquetti:2011fj,Beem:2012mb,Nieri:2013yra} \cite{Nedelin:2015mio, Nedelin:2016gwu, Cassia:2019sjk, Cassia:2020uxy}. 
See also a review article on this topic~\cite{Pasquetti:2016dyl}.
Based on this dictionary, some Virasoro limits were taken in the $q$-Virasoro CFT side that lead to matrix models which on the gauge theory side are associated with well-studied partition functions coming from supersymmetric localization.

Such correspondence was established by applying some techniques \cite{Mironov:1990im,Dijkgraaf:1990rs} to derive the Ward identities (which in the case of Virasoro CFTs we refer to as \textit{Virasoro constraints}). See also~\cite{Morozov:1994hh} for a review on this direction. These constraints are an (infinite) set of PDEs that are solved by the matrix model under study. When the matrix model under consideration is a Virasoro matrix model, these constraints can be written in the form of modes that satisfy Virasoro algebraic relation with a particular central charge.

On the other hand, one can move in the opposite direction. One can start with a Virasoro algebra with some choice of the central charge and then can follow an algorithm to construct the associated matrix model. In this algorithm one looks for an operator referred to as a \say{screening current} such that, if integrated along a particular choice of contour, the resulting operator, which we will refer to as the \say{screening charge} commutes with the Virasoro modes and hence it solves the Virasoro constraints.

The purpose of this paper is to extend the CFT description of the matrix model to a class of models that we call the \emph{Uglov matrix model}.
The partition function of degree $r$ is given by
\begin{align}
 \mathcal{Z}_\text{Uglov}(\underline{\tau};N) = \frac{1}{N!} \int \prod_{j=1}^N dx_j \, e^{-\frac{N\beta}{2} V(x_j)} \prod_{i<j}^N (x_i^r - x_j^r)^{2 (\beta - 1)/r} (x_i - x_j)^2
 \, ,
\end{align}
where we denote the potential function depending on the coupling constants $(\tau_n)_{n \in \mathbb{Z}_+}$ by $V(x)$.
See \eqref{uglov-matrix-model} for details.
The Uglov matrix model is obtained from the root of unity limit (Uglov limit) of the $\textbf{q}$-deformed matrix model,%
\footnote{%
As there exit two deformation parameters $(\textbf{q},\textbf{t})$, it is also possible to consider the root of unity limit only for either $\textbf{q}$ or $\textbf{t}$~\cite{Bouwknegt:1998da,Kimura:2019xzj}.
}
\begin{align}
 \textbf{q} \, ,  \textbf{t} \ \longrightarrow \ \exp \left( \frac{2 \pi i}{r} \right)
 \, .
\end{align}
Such a limit was originally considered to describe the spin generalization of the Calogero-Sutherland model by Uglov~\cite{Uglov:1997ia}.
See also a monograph on this topic \cite{Kuramoto:2009} for details.

As the Uglov matrix model is the root of unity limit of the $\textbf{q}$-matrix model, the corresponding CFT would be accordingly obtained from the root of unity limit of the $\textbf{q}$-Virasoro algebra.
For the ordinary limit of the $\textbf{q}$-Virasoro algebra, $\textbf{t} = \mathbf{q}^\beta$, $\textbf{q} \to 1$, it is known that the central charge of the corresponding CFT is given by~\cite{Shiraishi:1995rp}
\begin{align}
 c = 1 - 6 \mathcal{Q}_\beta^2
 \, , \qquad
 \mathcal{Q}_\beta = \sqrt{\beta} - \sqrt{\beta^{-1}}
 \, .
\end{align}
In this paper, we show that the central charge obtained from the Virasoro constraint of the Uglov matrix model of degree $r$ is given by
\begin{align}
 c = r - \frac{6 \mathcal{Q}_\beta^2}{r}
 \, .
 \label{eq:cc_deg_r}
\end{align}
This expression was conjectured in \cite{Kimura:2012sc} based on the generalization of the level-rank duality associated with the $\mathbb{Z}_r$ parafermion CFT~\cite{Fateev:1985mm,Fateev:1987zh}, which is motivated by the CFT description of the non-Abelian spin-singlet fractional quantum Hall wave function~\cite{Estienne:2012si}.
We remark that the central charge \eqref{eq:cc_deg_r} agrees with the result obtained through the correspondence between the 2d CFT and the 4d $\mathcal{N}=2$ gauge theory~\cite{Alday:2009aq,Wyllard:2009hg} on the asymptotically locally Euclidean (ALE) space given by the resolution of the orbifold denoted by $\widetilde{\mathbb{C}^2/\mathbb{Z}_r}$~\cite{Belavin:2011pp,Nishioka:2011jk}.
This agreement is naturally understood as the Uglov limit is also utilized to describe the 4d gauge theory on the orbifold $\widetilde{\mathbb{C}^2/\mathbb{Z}_r}$~\cite{Kimura:2011zf,Kimura:2011gq}.

The article is structured as follows. In Section~\ref{sec:qt_Virasoro}, we review the notion of $(\textbf{q},\textbf{t})$-deformed Virasoro algebra and its free field realization which we use to build the associated matrix model. Then, in Section~\ref{sec:U_limit}, we define the Uglov limit and we derive the matrix model associated with this limit (which we will refer to as \textit{Uglov matrix model}). We can do this following two paths: we either take this limit of the $\textbf{q}$-Virasoro matrix model or we take the limit at the algebra level and then derive the associated matrix model with that algebra and we end up with a matching between the two resulting models. 
Having the matrix model at hand, in Section~\ref{sec:V_const} we derive the associated Virasoro constraints which we can write in the form of some Virasoro modes. Studying the algebra of these modes we calculate the associated central charge either by direct computation or by using Sugawara construction. In Section~\ref{sec:Discussion} we give a  physical interpretation of our results and some further related work that can be done.

\section{(\textbf{q},\textbf{t})-Virasoro Partition Function}\label{sec:qt_Virasoro}
\subsection{$(\textbf{q},\textbf{t})$-deformed Virasoro algebra}
Take $\textbf{q},\textbf{t} \in \mathbb{C}^\times = \mathbb{C}\backslash \{0\}$. We start by introducing the notion of $(\textbf{q},\textbf{t})$\textit{-deformed Virasoro algebra} which we will denote by $\mathcal{V}ir_{\textbf{q},\textbf{t}}$. This algebra is generated by the modes $\{\textbf{T}_n \mid n \in \mathbb{Z}\}$ satisfying the relations~\cite{Shiraishi:1995rp}
\begin{equation}
\left[\textbf{T}_n, \textbf{T}_m\right] + \sum_{l\geq 1} \textbf{f}_l \left(\textbf{T}_{n-l}\textbf{T}_{m+l} - \textbf{T}_{m-l}\textbf{T}_{n+l}\right) = -\frac{(1-\textbf{q})(1-\textbf{t}^{-1})}{1-\textbf{p}} \left(\textbf{p}^n-\textbf{p}^{-n}\right) \delta_{n+m,0}.
\label{q,t-Virasoro-Algebra}
\end{equation}
The expansion coefficients $\textbf{f}_l$ are defined via the expansion
\begin{equation}
\sum_{l\geq 0} \textbf{f}_l x^l := \exp\left(\sum_{n>0} \frac{(1-\textbf{q}^n)(1-\textbf{t}^{-n})}{n (1+\textbf{p}^n)}x^n\right),
\end{equation}
with the new parameter $\textbf{p}$ defined as 
\begin{equation}
\textbf{p} = \textbf{q} \textbf{t}^{-1}.
\end{equation}

\subsubsection*{Free field realization}
One interesting representation for the algebra \eqref{q,t-Virasoro-Algebra} is the free boson representation. In this case we introduce the following $\mathcal{H}_{q,t}$ Heisenberg algebra 
\begin{equation}
\left[\textbf{a}_n, \textbf{a}_m\right] = \frac{1}{n} \left(\textbf{q}^{\frac{n}{2}} - \textbf{q}^{-\frac{n}{2}}\right)\left(\textbf{t}^{\frac{n}{2}} - \textbf{t}^{-\frac{n}{2}}\right)\left(\textbf{p}^{\frac{n}{2}} + \textbf{p}^{-\frac{n}{2}}\right)\delta_{n+m,0}, \quad \text{and}, \quad\left[\textbf{a}_n, \textbf{Q}\right] = \delta_{n,0}.
\label{Heisenberg-algebra}
\end{equation}
Taking the case where $\textbf{t} = \textbf{q}^\beta$, with $\beta \in \mathbb{C}$, the modes of the $\textbf{q}$-deformed Virasoro can be written in terms of the Heisenberg modes above as \cite{Shiraishi:1995rp}, 
\begin{equation}
\textbf{T}(x) := \sum_{n\in \mathbb{Z}} \textbf{T}_n x^{-n} = \textbf{p}^{-\frac{1}{2}}{:\exp\left(-\sum_{n\neq 0} \frac{x^{-n}}{(1+\textbf{p}^{n})}\textbf{a}_n\right):} \textbf{q}^{-\sqrt{\beta}\textbf{a}_0} +\textbf{p}^{\frac{1}{2}} {:\exp\left(\sum_{n\neq 0} \frac{x^{-n}}{(1+\textbf{p}^{-n})}\textbf{a}_n\right):} \textbf{q}^{\sqrt{\beta}\textbf{a}_0} .
\label{free-boson-realisation-of-q-virasoro-modes}
\end{equation}
By normal ordering we mean putting the negative oscillators to the right of the positive ones and \textit{the momentum operator} $\textbf{a}_0$ to the right of \textit{the charge operator} $\textbf{Q}$.

\subsubsection*{Scaling limit}
Being a $\textbf{q}$-deformation of the usual Virasoro algebra, one expects to recover the usual Virasoro modes in the limit where one takes $\textbf{q} = e^\hbar$ with $\hbar \rightarrow 0$. Applying this limiting case to the algebra \eqref{q,t-Virasoro-Algebra} with $\textbf{t}$ taking the particular form mentioned above, we end up with
\begin{equation}
	\textbf{T}_n = 2\delta_{n,0} + \hbar^2 \beta\left(\textbf{L}_n + \frac{1}{4}\mathcal{Q}_{\beta}^2\delta_{n,0} \right) + \mathcal{O}(\hbar^4),
\end{equation}
where
\begin{equation}
\mathcal{Q}_\beta := \sqrt{\beta} - \frac{1}{\sqrt{\beta}},
\end{equation}
and the generators $(\textbf{L}_n)_{n \in \mathbb{Z}}$ satisfy the Virasoro algebra
\begin{align}
    \left[\mathbf{L}_n, \mathbf{L}_m \right] = (n - m) \mathbf{L}_{n+m} + \frac{c}{12}(n^3 - n) \delta_{n+m,0}
    \, ,
\end{align}
with central charge given by
\begin{equation}
c = 1 - 6 \mathcal{Q}_\beta^2.
\end{equation}

\subsubsection*{Screening current}
Going back to $\textbf{q}$-Virasoro, we define a \textit{screening-current} $\mathcal{S}(x)$ to be such that
\begin{equation}
	\left[\textbf{T}(x), \mathcal{S}(x)\right] = \frac{\textbf{O}(\textbf{q}x) - \textbf{O}(x)}{x},
	\label{definition-of-screening-current}
\end{equation}
with $\textbf{O}(x)$ being some function. It has been found \cite{Shiraishi:1995rp} that there are two operators satisfying the this definition, 
\begin{subequations}
\begin{equation}
\mathcal{S}_{\textbf{q}}(x):= {:\exp\left(-\sum_{n\neq 0} \frac{x^{-n}}{\textbf{q}^{\frac{n}{2}}-\textbf{q}^{-\frac{n}{2}}}\textbf{a}_n\right):} \exp\left(\sqrt{\beta}\textbf{Q}\right)x^{2\sqrt{\beta}\textbf{a}_0},
\label{q-screening-current}
\end{equation}
\begin{equation}
\mathcal{S}_{\textbf{t}}(x):= {:\exp\left(-\sum_{n\neq 0} \frac{x^{-n}}{\textbf{t}^{-\frac{n}{2}}-\textbf{t}^{\frac{n}{2}}}\textbf{a}_n\right):} \exp\left(-\frac{\textbf{Q}}{\sqrt{\beta}}\right)x^{-\frac{2\textbf{a}_0}{\sqrt{\beta}}}.
\end{equation}
\end{subequations}
We can see that there is a similarity between these two solutions. Hence, we keep going with our discussion using only the first one.%
\footnote{%
By considering these two screening currents at the same time, one can discuss the supergroup analog of the ($\textbf{q}$-)matrix model~\cite{Kimura:2021ngu,Cassia:2021uly}.
}

The reason behind referring to such operators as \say{currents} is that one can pick a particular contour $\mathcal{C}$ which is a Fleder's cycle along which the integral on the right-hand side of \eqref{definition-of-screening-current} vanishes. With this observation we define \textit{the associated screening charge} to be the integral of the current along that cycle
\begin{equation}
\mathcal{J}_{\textbf{q}} := \oint_{\mathcal{C}} \frac{dx}{2\pi \textbf{i}} \mathcal{S}_{\textbf{q}}(x).
\end{equation}

\subsection{$\mathbf{q}$-Virasoro matrix model}

As the last object to introduce in this section we define the $\textbf{q}$\textit{-Virasoro matrix model}, denoted by $\mathcal{Z}_{\textbf{q}}$ to be the solution of $\textbf{q}$-Virasoro constraints \cite{Awata:1994xd,Nedelin:2015mio,Nedelin:2016gwu,Lodin:2018lbz,Cassia:2019sjk,Cassia:2020uxy}
\begin{equation}
\textbf{T}_{n>0} \mathcal{Z}_{\textbf{q}} = 0.
\end{equation}
With the construction of the screening currents we made above we note that these equations can be solved by taking
\begin{equation}
\mathcal{Z}_{\textbf{q}} = \mathcal{J}_{\textbf{q}}^N,
\label{state-form-of-matrix-model-q-virasoro}
\end{equation}
for some positive integer $N$.

To see the matrix model form more explicitly we first need to note that if we take $\mathcal{S}_{\textbf{q}}^+(x)$ ($\mathcal{S}_\textbf{q}^{-}(x)$) to be the part of the exponential containing the positive (negative) Heisenberg oscillators in \eqref{q-screening-current}, then, we have 
\begin{equation}
	 \mathcal{S}_{\textbf{q}}^+(x_i) \mathcal{S}_{\textbf{q}}^- (x_j) =  \mathcal{S}_{\textbf{q}}^-(x_j)\mathcal{S}_{\textbf{q}}^+(x_i) \frac{\left(x_j x_i^{-1};\textbf{q}\right)_\infty \left( \textbf{q}\textbf{t}^{-1}x_jx_i^{-1};\textbf{q}\right)_\infty}{\left( \textbf{t}x_jx_i^{-1};\textbf{q}\right)_\infty\left(\textbf{q}x_jx_i^{-1};\textbf{q}\right)_\infty}.
\end{equation}
Here we used the $\textbf{q}$\textit{-Pochhammer symbol}
\begin{equation}
\left(x;\textbf{q}\right)_\infty := \prod_{k\geq 0} \left(1 - \textbf{q}^k x\right).
\end{equation}

With this observation one can deduce the following OPE for the $\textbf{q}$-screening currents 
\begin{equation}
	\prod_{j=1}^N \mathcal{S}_{\textbf{q}}(x_j) = {:\prod_{j=1}^N \mathcal{S}_{\textbf{q}}(x_j):} \Delta_{\beta}(\underline{x}; \textbf{q})\textbf{c}_\beta(\underline{x}; \textbf{q}) \prod_{j=1}^N x_j^{\beta(N-1)},
\label{q-virasoro-currents-ope}
\end{equation}
where $\Delta_{\beta}(\underline{x};\textbf{q})$ is the $\textbf{q}$\textit{-Vandermonde determinant} 
\begin{equation}
 \Delta_\beta (\underline{x};\textbf{q}) := \prod_{k\neq j}^N \frac{(x_k x_j^{-1}; \textbf{q})_{\infty}}{(\textbf{t}x_kx_j^{-1};\textbf{q})_{\infty}}
\end{equation}
Meanwhile, the other factor is given by
\begin{equation}
\textbf{c}_\beta (\underline{x};\textbf{q}):= \prod_{k<j} (x_kx_j^{-1})^\beta \frac{\Theta(\textbf{t}x_k x_j^{-1};\textbf{q})}{\Theta(x_kx_j^{-1};\textbf{q})}.
\end{equation}
Here we also introduced the $\Theta$-function defined in terms of $\textbf{q}$\textit{-Pochhammer symbol} as
\begin{equation}
	\Theta(x;\textbf{q}) := (x;\textbf{q})_\infty \left(\textbf{q}x^{-1};\textbf{q}\right)_\infty
\end{equation}

Going back to the Heisenberg algebra $\mathcal{H}_{q,t}$ in \eqref{Heisenberg-algebra} we define the vacuum state to be annihilated by all the positive oscillators along with the momentum operator. Moreover, due to the non-vanishing commutator of the momentum with the charge operator, one can construct charged-vacuum states which are momentum eigenstates:
\begin{equation}
	\left| \alpha \right> := \exp\left(\alpha \textbf{Q}\right)\left| 0 \right>, \quad \textbf{a}_0\left|\alpha\right> = \alpha \left|\alpha\right>.
\end{equation}
An interesting point about this observation is that now we can pick a representation of the algebra \eqref{Heisenberg-algebra} such that we can explicitly write down the matrix model form of $\mathcal{Z}_{\textbf{q}}$ defined in \eqref{state-form-of-matrix-model-q-virasoro}. In order to accomplish this, let's consider the following representation in terms of the \textit{time variables} $\underline{\tau} := (\tau_0, \tau_1, \tau_2, \cdots)$ \cite{Nedelin:2016gwu}
\begin{equation} 
	\textbf{a}_{-n>0} \sim \left(\textbf{q}^{\frac{n}{2}} - \textbf{q}^{-\frac{n}{2}}\right)\tau_{n} , \quad \textbf{a}_{n>0} \sim \frac{1}{n} \left(\textbf{t}^{\frac{n}{2}} - \textbf{t}^{-\frac{n}{2}}\right)\left(\textbf{p}^{\frac{n}{2}} + \textbf{p}^{-\frac{n}{2}}\right)\frac{\partial}{\partial\tau_n}.
\end{equation}
As for the momentum and charge operators,
\begin{equation}
\textbf{Q} \sim \sqrt{\beta}\tau_0, \quad \textbf{a}_0\sim\frac{1}{\sqrt{\beta}} \frac{\partial}{\partial\tau_0}, \quad 
\text{and,} \quad \left|\alpha\right> \sim \exp\left(\sqrt{\beta}\tau_0\right)\cdot 1.
\end{equation}
With respect to the representation, the matrix model associated with $\textbf{q}$-Virasoro algebra is given by,
\begin{equation}
	\mathcal{Z}_{\textbf{q}} \left|\alpha\right> \sim \mathcal{Z}_{\textbf{q}} (\underline{\tau}) =: \mathcal{N}_{\textbf{q}} \oint_{\cup_{j=1}^N \mathcal{C}_j} \prod_{j=1}^N \frac{dx_j}{2\pi \textbf{i}} \Delta_{\beta}(\underline{x};\textbf{q}) \textbf{c}_\beta (\underline{x};\textbf{q}) \exp\left(\sum_{j=1}^N V_{\textbf{q}} \left(x_j\mid \underline{\tau}\right)\right),
	\label{q-Virasoro-Matrix-Model}
\end{equation}
where the normalization factor and the potential function are given by
\begin{subequations}
\begin{align}
\mathcal{N}_{\textbf{q}} &:= \exp\left(\kappa_0 \tau_0\right),\\
V_{\textbf{q}} (x\mid\underline{\tau}) &:= \kappa_1 \ln x + \sum_{n>0} \tau_n x^n, 
\label{q-Virasoro-potential}
\end{align}
\end{subequations}
with
\begin{equation}
\kappa_0 := N + {\alpha}\sqrt{\beta}, \quad \text{and,} \quad \kappa_1 := \sqrt{\beta}\left(\alpha +\sqrt{\beta}N-\mathcal{Q}_\beta\right).
\end{equation}

The measure of this matrix model is the same as that of the so-called 3d holomorphic blocks \cite{Pasquetti:2011fj,Beem:2012mb,Nieri:2013yra} coming from localizing the partition function associated with $\mathcal{N}=2$ $U(N)$ supersymmetric Chern-Simon-Yang-Mills theory on $\mathbb{D}^2\times \mathbb{S}^1$ with a matter multiplet in the adjoint representation~\cite{Yoshida:2014ssa}. With this observation \cite{Nedelin:2016gwu} a correspondence can be established between 3d $\mathcal{N}=2$ Chern-Simons theory and 2d $\textbf{q}$-Virasoro CFT where the parameters $\kappa_0$ and $\kappa_1$ defined above are translated in terms of the Chern-Simons level and FI coupling constants of the 3d theory.

\section{Uglov Limit of $\textbf{q}$-Virasoro Matrix Model}
\label{sec:U_limit}

As we saw earlier, one can get the usual Virasoro algebra from the $\textbf{q}$-deformed algebra \eqref{q,t-Virasoro-Algebra} by taking the $\textbf{q}\rightarrow 1$ limit, $\hbar \rightarrow 0$ with $\textbf{q} = e^\hbar$. 
In this case, the central charge of the Virasoro algebra is given by $c = 1 - 6\mathcal{Q}_\beta^2$. 
Another way to arrive to this result can be accomplished by starting with the $\textbf{q}$-deformed matrix model \eqref{q-Virasoro-Matrix-Model} and study the behaviour of its measure under the limit $\textbf{q} \rightarrow 1$. We will carry out this procedure more explicitly later for the so-called \textit{Uglov limit} which will be defined momentarily. Meanwhile, for the case at hand it is found that the resulting matrix model is nothing other than the well-known $\beta$-ensemble model which is a Virasoro matrix model associated with the central charge mentioned above.

To see more explicitly how these techniques work, we take the special limit that we mentioned above: \textit{Uglov limit}~\cite{Uglov:1997ia}.
This is a root-of-unity limit of the deformation parameter $\textbf{q}$ where we take 
\begin{equation}
	\textbf{q} \rightarrow \omega \textbf{q}, \quad \textbf{t} \rightarrow \omega \textbf{q}^\beta, \quad \text{and} \quad \textbf{q} \rightarrow 1,
	\label{Uglov-limit}
\end{equation}
where $\omega  := \exp\left(\frac{2\pi \textbf{i}}{r}\right)$, for some $r \in \mathbb{Z}_{+}$ which we will refer to as \textit{the order of Uglov limit}.
In this Section, we study the Uglov matrix model defined by the Uglov limit of the $\textbf{q}$-deformed matrix model.
In particular, we explore the Virasoro constraint associated with the Uglov matrix model to discuss the underlying CFT model.

\subsection{Limit of $(\textbf{q},\textbf{t})$-matrix model}

We consider the case $\beta = kr + 1 \in r\mathbb{Z}_+ + 1$ for the moment. Using the identity,
\begin{equation}
	\prod_{n=1}^{kr} \left(1-\omega^n x\right) = \left(1-x^r\right)^k,
\end{equation}
one can see that, under the limit \eqref{Uglov-limit}, the Vandermonde-like part of the measure reduces to the following form\footnote{Here we define the Uglov measure up to a \say{zero mode} factor $\prod_{i\neq j} x_i^{-\beta}$ that we are not going to worry about here.}
\begin{equation}
	\Delta_{\beta}\left(\underline{x}; \textbf{q}\right)\Big|_{(\textbf{q}, \textbf{t}) \rightarrow (\omega \textbf{q}, \omega \textbf{q}^\beta)} = \prod_{i\neq j}^N \frac{\left(x_ix_j^{-1};\omega\textbf{q}\right)_\infty}{\left(\omega \textbf{q}^\beta x_ix_j^{-1};\omega \textbf{q}\right)_\infty} \xrightarrow[{\beta = kr+1}]{\textbf{q}\rightarrow 1}  \Delta^{(r)}_\text{Uglov}(\underline{x}) :=\prod_{i<j}^N \left(x_i^r - x_j^r\right)^{2k} \left(x_i - x_j\right)^2.
	\label{uglov-measure-factor}
\end{equation}
We emphasize that this expression is also available for arbitrary $\beta \in \mathbb{C}$~\cite{Kimura:2011gq}.
As for the second part of the measure, repeating the same steps one can see that this part reduces to a numerical value which does not affect the matrix model.

As we mentioned before, the momentum eigenstate of the charge vacuum $\alpha$ is a free parameter that we picked to make the connection between the $\textbf{q}$-Virasoro matrix model and supersymmetric Chern-Simons partition function manifest. With that begin said, it is still a free parameter that the central charge associated with the resulting Virasoro model should not depend on. Hence, we will take the uncharged-vacuum: $\alpha = 0$.

A special point about this value is that  the logarithmic part of the potential \eqref{q-Virasoro-potential} gets cancelled out with a similar form coming from the measure and we can redefine the time parameters $\underline{\tau}$ such that we end up with the following model
\begin{equation}
	\mathcal{Z}_\text{Uglov} (\underline{\tau}; N) := \oint \prod_{j=1}^N dx_j \textbf{Z}_{N}(\underline{x}\mid \underline{\tau}), 
	\label{uglov-matrix-model}
\end{equation}
where, 
\begin{equation}	
 \textbf{Z}_N(\underline{x}\mid\underline{\tau}) := \frac{1}{N!} \prod_{i<j}^N \left(x_i^r - x_j^r\right)^{2k} \left(x_i - x_j\right)^2\exp\left(-\frac{N\beta}{2} \sum_{j=1}^N \sum_{n>0}\frac{\tau_n}{n} x_j^n\right).
\end{equation}
We will refer to this as \textit{Uglov matrix model of degree $r$}.

Now that we have the matrix model we will now move to discussing the associated Virasoro constraint.
We will find that the central charge matches the form conjectured in \cite{Kimura:2012sc} which is the main result in this article. But before we do that, let us first give another way of deriving the model \eqref{uglov-matrix-model} where instead of applying the Uglov limit \eqref{Uglov-limit} on the $\textbf{q}$-Virasoro matrix model \eqref{q-Virasoro-Matrix-Model} we apply it on  $\textbf{q}$-Virasoro algebra \eqref{q,t-Virasoro-Algebra}.

\subsection{Limit of screening current}

In the Uglov limit \eqref{Uglov-limit}, the ($\mathbf{q}$,$\mathbf{t}$)-deformed algebra \eqref{Heisenberg-algebra} reduces to the mod $r$ Heisenberg algebra~\cite{Itoyama:2013mca,Itoyama:2014pca}
\begin{subequations}\label{Uglov-algebra}
\begin{align}
\left[\textbf{a}_n, \textbf{Q}\right] &= \delta_{n,0},\\
\left[\textbf{a}_n, \textbf{a}_m\right] &= n\delta_{n+m,0},\\
\left[\tilde{\textbf{a}}_{n+\frac{\ell}{r}}, \tilde{\textbf{a}}_{-m-\frac{\ell'}{r}}\right] &= \left(n + \frac{l}{r}\right) \delta_{n,m} \delta_{\ell, \ell'}. 
\end{align}
\end{subequations}
As for the $\textbf{q}$-Screening current \eqref{q-screening-current}, in this limit it becomes of the form,\footnote{The normal ordering here is same as defined earlier.}
\begin{equation}
	\mathcal{S}_\text{Uglov} (x) := x^{\frac{\beta}{r}} \exp\left(\sqrt{\frac{2\beta}{r}}\textbf{Q}\right)x^{\sqrt{\frac{2\beta}{r}}\textbf{a}_0} {\colon\mathcal{A}_{r} (x\mid \underline{\textbf{a}})\colon}  {\colon\mathcal{B}_{r}(x\mid\underline{\tilde{\textbf{a}}})\colon},
\end{equation}
where
\begin{subequations}
\begin{align}
	\mathcal{A}_r(x\mid\underline{\textbf{a}}) &:= \exp\left(-\sqrt{\frac{2\beta}{r}}\sum_{n\neq 0} \frac{x^{-n}}{n}\textbf{a}_{n}\right), \\
	 \mathcal{B}_r(x\mid\underline{\tilde{\textbf{a}}}) &:= \exp\left(\sqrt{\frac{2}{r}}\sum_{\ell=1}^{r-1}\sum_{n\in \mathbb{Z}} \frac{x^{-n-\frac{\ell}{r}}}{n+\frac{\ell}{r}} \tilde{\textbf{a}}_{n+\frac{\ell}{r}}\right).
\end{align}
\end{subequations}

In order to derive the OPEs of the above screening currents, we first observe that the algebra \eqref{Uglov-algebra} leads to the following OPE
\begin{equation}
\mathcal{A}_r^{(+)}(x_i\mid\underline{\textbf{a}}) \mathcal{A}_{r}^{(-)}(x_j\mid\underline{\textbf{a}}) = \left(1 - \frac{x_j}{x_i}\right)^{\frac{2\beta}{r}} \mathcal{A}_{r}^{(-)}(x_j\mid\underline{\textbf{a}}) \mathcal{A}_r^{(+)}(x_i\mid\underline{\textbf{a}}),
\end{equation}
with $\mathcal{A}^{(+)}_{r}$($\mathcal{A}^{(-)}_{r}$) standing for the component of $\mathcal{A}_r$ with the positive (negative) oscillators in the argument of the exponential.
For $N \in \mathbb{Z}_+$, Uglov screening-currents, this observation can be generalized into 
\begin{equation}
	\prod_{i=1}^N {\colon\mathcal{A}_r(x_i\mid\underline{\textbf{a}})\colon}  = \prod_{1\leq i<j\leq N} \left(1-\frac{x_j}{x_i}\right)^{\frac{2\beta}{r}} {\colon\prod_{i=1}^{N} \mathcal{A}_r(x_i\mid\underline{\textbf{a}})\colon}.
\end{equation}

Moreover, observing that
\begin{equation}
\sum_{\ell=1}^{r-1}\sum_{n>0} \frac{x^{n+\frac{\ell}{r}}}{n+\frac{\ell}{r}} = \log\frac{1-x}{\left(1-x^{\frac{1}{r}}\right)^r},
\end{equation}
one can compute the OPE coming from the $\mathcal{B}_r$-factor of the Uglov current to be 
\begin{equation}
	\mathcal{B}_r^{(+)}(x_i\mid\underline{\tilde{\textbf{a}}}) \mathcal{B}_r^{(-)}(x_j\mid\underline{\tilde{\textbf{a}}}) = \left(1- \frac{x_j^{\frac{1}{r}}}{x_i^{\frac{1}{r}}}\right)^2\left(1-\frac{x_j}{x_i}\right)^{\frac{2}{r}}\mathcal{B}_r^{(-)}(x_j\mid\underline{\tilde{\textbf{a}}}) \mathcal{B}_r^{(+)}(x_i\mid\underline{\tilde{\textbf{a}}}).
\end{equation}
This can be generalized for the case of $N \in \mathbb{Z}_+$ currents as
\begin{equation}
	\prod_{i=1}^N {\colon\mathcal{B}_r(x_i\mid\underline{\tilde{\textbf{a}}})\colon} = \prod_{1\leq i<j\leq N} \left(1- \frac{x_j^{\frac{1}{r}}}{x_i^{\frac{1}{r}}}\right)^2\left(1-\frac{x_j}{x_i}\right)^{\frac{2}{r}} \cdot {\colon\prod_{i=1}^N \mathcal{B}_r(x_i\mid\underline{\tilde{\textbf{a}}})\colon}.
\end{equation}

Putting these results together along with the factor resulting from the non-vanishing commutator of the $\textbf{a}_0$ with the charge operator $\textbf{Q}$, we end up with the following OPE
\begin{equation}
	\prod_{i=1}^N \mathcal{S}_\text{Uglov}(x_i) =  \mathcal{D}^{(r)}_{\beta}(\underline{x}) {:\prod_{i=1}^N \mathcal{S}_\text{Uglov}(x_i):},
\end{equation}
with the OPE factor 
\begin{equation}
	\mathcal{D}_{\beta}^{(r)}(\underline{x}) := \prod_{1\leq i<j\leq N} \left(x_i^{\frac{1}{r}}-x_j^{\frac{1}{r}}\right)^2\left(x_i-x_j\right)^{2\frac{\beta-1}{r}} = \Delta_\text{Uglov}^{(r)} \left(\underline{x}^{\frac{1}{r}}\right),
\end{equation}
where from the last equality we can see that matching between this measure with that of Uglov matrix model we derived in \eqref{uglov-measure-factor}.

\section{Virasoro Constraint of Uglov Matrix Model}
\label{sec:V_const}

Going back to the model \eqref{uglov-matrix-model} we now come to extracting what Virasoro algebra is it associated with. Following \cite{Mironov:1990im,Dijkgraaf:1990rs} we will work out the explicit form of the Virasoro modes where, as we shall see, this can be explicitly done for a particular subset of these modes.

\subsubsection*{Free field realization}

To do this, let us start with the observation that the time-derivative of the integrand of \eqref{uglov-matrix-model} yields the power sum symmetric polynomial 
\begin{equation}
	p_n(\underline{x}) := \sum_{j=1}^N x_j^n = -\frac{2n}{N\beta} \frac{\partial}{\partial\tau_n} \log \textbf{Z}_N (\underline{x}\mid\underline{\tau}).
\end{equation}
Taking the following representation for $n>0$
\begin{equation}
	\textbf{a}_n \sim \frac{2n}{N\sqrt{\beta}} \frac{\partial}{\partial\tau_n}, \quad \textbf{a}_{-n} \sim \frac{N\sqrt{\beta}}{2}\tau_n, 
	\label{representation-of-uglov-algebra}
\end{equation}
one can see that these modes are Heisenberg oscillators in the sense that 
\begin{equation}
[\textbf{a}_{n}, \textbf{a}_{m}] = n\delta_{n+m, 0}.
\label{uglob-heisenberg-algebra}
\end{equation}
This implies that the integrand of Uglov model is an eigenstate of the positive-Heisenberg oscillators 
\begin{equation}
p_n(\underline{x}) \textbf{Z}_{N}(\underline{x}\mid\underline{\tau}) = -\frac{1}{\sqrt{\beta}}\textbf{a}_n \textbf{Z}_{N}(\underline{x}\mid\underline{\tau}).
\end{equation}

\subsubsection*{Ward identity}

The Virasoro constraints are nothing other than the Ward identities resulting from the invariance of our model \eqref{uglov-matrix-model} under the shift 
\begin{equation}
\delta_n x_i := \epsilon_n x_i^{n+1}.
\end{equation}
Under this shift, the Vandermonde-like part of the measure changes as 
\begin{equation}
\delta_n \Delta_\text{Uglov}^{(r)}(\underline{x}) = \Delta_\text{Uglov}^{(r)}(\underline{x}) \cdot \left(2\sum_{i<j} \frac{x_i^{n+1} - x_j^{n+1}}{x_i - x_j} + 2kr\sum_{i<j} \frac{x_i^{r+n} -x_j^{r+n}}{x_i^r - x_j^r}\right).
\label{variation-of-vandermonde-like-measure}
\end{equation}
Meanwhile, the other part of the measure changes as
\begin{equation}
\delta_n \prod_{j=1}^N dx_j = \epsilon_n (n+1) \sum_{j=1}^N x_j^n \cdot \prod_{j=1}^{N} dx_j = \epsilon_n (n+1) p_n(\underline{x}) \cdot \prod_{j=1}^N dx_j.
\end{equation}
As for the argument of the exponential, 
\begin{equation}
\delta_n \left(-\frac{N\beta}{2} \sum_{m>0} \frac{\tau_m}{m} p_m(\underline{x})\right) = -\epsilon_n \frac{N\beta}{2} \sum_{m>0} \tau_m p_{m+n}(\underline{x}).
\end{equation}

To re-write \eqref{variation-of-vandermonde-like-measure} in terms of the polynomials $p_n(\underline{x})$ we note the identity 
\begin{equation}
	\sum_{i\neq j}^N \frac{x_i^{n+1} - x_j^{n+1}}{x_i - x_j} = \sum_{m=0}^n p_m(\underline{x}) p_{n-m}(\underline{x}) - (n+1) p_n(\underline{x}).
\end{equation}
This identity works for positive modes $n$. At the time of writing this article we did not have a more generic identity that we could apply for the the second sum in \eqref{variation-of-vandermonde-like-measure} so we are restricting to the case where $n = r\tilde{n}$ for some positive $\tilde{n}$.%
\footnote{%
The Uglov matrix model is originally introduced to describe the $\mathbb{Z}_r$ orbifold theory~\cite{Kimura:2011zf,Kimura:2011gq}.
From this point of view, the mode $n \not\in r\mathbb{Z}$ is not $\mathbb{Z}_r$ invariant, so that we do not consider such a case.
}
In this case, noting that $p_n(\underline{x^r}) = p_{rn}(\underline{x})$, we can see that the second sum can be written as
\begin{equation}
	\sum_{1\leq i\neq j\leq N} \frac{x_i^{r\tilde{n}+r} - x_j^{r\tilde{n}+r}}{x_i^r - x_j^r} = \sum_{m=0}^{\tilde{n}} p_{rm} (\underline{x}) p_{r\tilde{n}-rm}(\underline{x}) - (\tilde{n}+1) p_{r\tilde{n}}(\underline{x}).
 \end{equation}
 
Putting these results together, the corresponding Ward identity is
 \begin{multline}
\delta_{rn}\mathcal{Z}_\text{Uglov} (\underline{\tau}; N)= 0 \Rightarrow \\ \left<-\frac{N\beta}{2}\sum_{m>0} \tau_m p_{m+rn}(\underline{x}) -kr(n+1)p_{rn}(\underline{x}) +kr\sum_{m=0}^n p_{rm}(\underline{x}) p_{rn-rm}(\underline{x}) + \sum_{m=0}^{rn} p_m(\underline{x}) p_{rn-m}(\underline{x})\right> =0,
 \end{multline}
 which, using the representation \eqref{representation-of-uglov-algebra}, can be re-written as the form of the Virasoro constraint
 \begin{equation}
 	\textbf{L}_{rn} \mathcal{Z}_\text{Uglov}(\underline{\tau};N) = 0
 	\quad \text{for} \quad n \ge 0
 \end{equation}
 The differential operator is now given by
  \begin{equation}
 \textbf{L}_{rn} := \sum_{m>0} \textbf{a}_{-m} \textbf{a}_{m+rn} +\frac{1}{\beta}\sum_{m=0}^{rn} \textbf{a}_m \textbf{a}_{rn-m} + \frac{\mathcal{Q}_\beta}{\sqrt{\beta}} \sum_{m=0}^{n} \textbf{a}_{rm}\textbf{a}_{rn-rm} + \mathcal{Q}_\beta(n+1)\textbf{a}_{rn}.
 \label{uglov-modes}
 \end{equation}

\subsubsection*{Virasoro generator of degree $r$}

We have the decomposition
\begin{align}
    \sum_{m=0}^{rn} \textbf{a}_m \textbf{a}_{rn-m}
    = \sum_{m=0}^{n} \textbf{a}_{rm} \textbf{a}_{rn-rm} + \sum_{\ell=1}^{r-1} \sum_{m=0}^{n-1} \textbf{a}_{rm+\ell} \textbf{a}_{rn-rm-\ell }
    \, ,
\end{align}
which, along with the redefinition $(\textbf{a}_{-n>0}, \textbf{a}_{n>0})\mapsto (\sqrt{2}\textbf{a}_{-n>0}, 1/\sqrt{2} \textbf{a}_{n>0})$, yields the following expression of the modes \eqref{uglov-modes} 
\begin{align} 
    \textbf{L}_{rn} & = 
    \sum_{m>0} \textbf{a}_{-m} \textbf{a}_{rn+m} +  \frac{1}{2}\sum_{m=0}^{n} \textbf{a}_{rm} \textbf{a}_{rn-rm} + \frac{\mathcal{Q}_\beta}{\sqrt{2}} (n+1) \textbf{a}_{rn}
    \nonumber \\ 
    & \qquad + 
   \sum_{\ell=1}^{r-1} \left[ \sum_{m>0} \textbf{a}_{-rm-\ell } \textbf{a}_{rn + rm + \ell} +   \frac{1}{2} \sum_{m=0}^{n-1} \textbf{a}_{rm+\ell } \textbf{a}_{rn-rm-\ell} \right]
    \, .
    \label{eq:LRN_op}
\end{align}
We see that it obeys the Virasoro algebraic relation,
\begin{align}
    \left[ \mathbf{L}_{rn}, \mathbf{L}_{rn'} \right] = r (n - n') \mathbf{L}_{r(n+n')}
    \, , \qquad n, n' \ge 0 \, .
\end{align}

We define a new operator for $n \in \mathbb{Z}_{\ge 0}$,
\begin{align}
    L_{n}^{(r,\ell)} & = 
    \begin{cases}
    \displaystyle
    \frac{1}{r}
    \sum_{m>1} \textbf{a}_{-rm} \textbf{a}_{rn+rm} + \frac{1}{2r} \sum_{m=0}^{n} \textbf{a}_{rm} \textbf{a}_{rn-rm} + \frac{\mathcal{Q}_\beta}{\sqrt{2}r} (n+1) \textbf{a}_{rn}
    & (\ell = 0) \\ \displaystyle
    \frac{1}{r}
    \sum_{m>0} \textbf{a}_{-rm-\ell } \textbf{a}_{rn + rm + \ell} + \frac{1}{2r} \sum_{m=0}^{n-1} \textbf{a}_{rm+\ell } \textbf{a}_{rn-rm-\ell }
    & (\ell \neq 0)
    \end{cases}
    \, ,
\end{align}
which is related to the original one \eqref{eq:LRN_op} as follows,
\begin{align}
    \textbf{L}_{rn} = \sum_{\ell=0}^{r-1} r L^{(r,\ell)}_{n}
    \, .
\end{align}
We see that they obey the Virasoro algebraic relation,
\begin{align}
    \left[L^{(r,\ell)}_n, L^{(r,\ell')}_{n'}\right] = (n-n') \, L^{(r,\ell)}_{n+n'} \, \delta_{\ell,\ell '}
    \, , \qquad
    n, n' \ge 0
    \, .
\end{align}

We can analytically continue this result for $n \in \mathbb{Z}$ by applying the following definition,
\begin{align}
    \textbf{L}_{n}^{(r,\ell)} & = 
    \begin{cases}
    \displaystyle
    \frac{1}{2r}
    \sum_{m \in \mathbb{Z}} {: \textbf{a}_{-rm} \textbf{a}_{rn+rm} :} + \frac{\mathcal{Q}_\beta}{\sqrt{2}r} (n+1) \textbf{a}_{rn}
    & (\ell = 0) \\ \displaystyle
    \frac{1}{2r}
    \sum_{m\in \mathbb{Z}} {:\textbf{a}_{-rm-\ell } \textbf{a}_{rn + rm + \ell} :} 
    & (\ell = r/2, r \in 2 \mathbb{N})
    \\ \displaystyle
    \frac{1}{r}
    \sum_{m\in \mathbb{Z}} {:\textbf{a}_{-rm-\ell } \textbf{a}_{rn + rm + \ell} :} 
    & (\text{otherwise}).
    \end{cases}
    \, .
    \label{eq:Lk-1_op_gen}
\end{align}
One important relation to remark is
\begin{align}
    \textbf{L}_n^{(r,\ell)} = \textbf{L}_n^{(r,r-\ell )}
    \, , \qquad
    n \in \mathbb{Z}
    \, ,
\end{align}
and thus the generator $\textbf{L}_n^{(r,\ell)}$ is decomposed into ${L}_n^{(r,\ell)}$ and ${L}_n^{(r,r-\ell)}$ for $\ell \not\in 0, r/2$ in the previous convention.
With this in mind, one can see that the modes \eqref{eq:Lk-1_op_gen} satisfy the following Virasoro algebra,
\begin{align}
    \left[\textbf{L}^{(r,\ell)}_n,\textbf{L}^{(r,\ell')}_{n'}\right] = (n-n') \, \textbf{L}^{(r,\ell)}_{n+n'} \, \delta_{\ell,\ell '} + \frac{c_{r,\ell}}{12} (n^3 - n) \delta_{n+n',0} \delta_{\ell,\ell '}
    \, ,
\end{align}
with the central charge
\begin{align}
    c_{r,\ell} =
    \begin{cases}
    \displaystyle
    1 - \frac{6 \mathcal{Q}_\beta^2}{r} & (\ell=0) \\
    1 & (\ell = r/2, r \in 2\mathbb{N}) \\
    2 & (\text{otherwise})
    \end{cases}
    \, .
    \label{eq:cc}
\end{align}
This result is concisely obtained from the OPE of the corresponding current operators as shown below.
As for the total central charge, we have
\begin{align}
    c_r 
    =
    \sum_{\ell = 0}^{\lfloor r/2 \rfloor}
    c_{r,\ell}
    = r - \frac{6 \mathcal{Q}_\beta^2}{r}
    \, ,
\end{align}
which agrees with the central charge associated with the Uglov limit of the $\textbf{q}$-Virasoro algebra conjectured in \cite{Kimura:2012sc}.

\subsubsection*{Sugawara construction}

We have the Sugawara construction of the Virasoro generators of degree $r$ with the following \textit{degree $r$ current},\footnote{Namely, one can define the current $\textbf{J}_{r,\ell}(x)$ in terms of the mod $r$ Heisenberg algebra~\eqref{Uglov-algebra}.}
\begin{align}
    \textbf{J}_{r,\ell}(x) = \sum_{m \in \mathbb{Z}} \frac{\textbf{a}_{rm+\ell }}{x^{m+1+\ell /r}}
    \, .
\end{align}

Using the algebra \eqref{uglob-heisenberg-algebra}, we can calculate the following OPE,
\begin{align}
    \textbf{J}_{r,\ell}(x) \textbf{J}_{r,\ell'}(y) & = \sum_{m, m' \in \mathbb{Z}} \frac{\textbf{a}_{rm+\ell }}{x^{m+1+\ell /r}} \frac{\textbf{a}_{rm'+\ell '}}{y^{m'+1+\ell '/r}}
    \nonumber \\
    & = {: \textbf{J}_{r,\ell}(x) \textbf{J}_{r,\ell'}(y) :} +
    \left(\frac{y}{x}\right)^{\ell/r} \frac{r }{(x - y)^2} \delta_{\ell+\ell ',0}+ \ell \left(\frac{y}{x}\right)^{\ell/r} \frac{x^{-1}}{x-y} \delta_{\ell+\ell ',0}
    \, .
\end{align}
Recalling
\begin{subequations}
\begin{align}
    \left(\frac{y}{x}\right)^{l/r} \frac{r}{(x - y)^2}
    & = \frac{r}{(x - y)^2} - \ell \frac{y^{-1}}{x - y} + \cdots
    \, , \\
    \ell \left(\frac{y}{x}\right)^{l/r} \frac{x^{-1}}{x-y}
    & = \ell \frac{y^{-1}}{x - y} + \cdots
    \, ,
\end{align}
\end{subequations}
we obtain the following OPEs for the degree $r$ currents,
\begin{subequations}\label{eq:JJ_OPE}
\begin{align}
    \textbf{J}_{r,\ell}(x) \textbf{J}_{r,\ell'}(y) & = \frac{r}{(x - y)^2}\delta_{\ell+\ell ',0} + \cdots \, , \\
    \partial_x \textbf{J}_{r,\ell}(x) \textbf{J}_{r,\ell'}(y) & = -\frac{2r }{(x - y)^3}\delta_{\ell+\ell ',0} + \cdots \, , \\
    \textbf{J}_{r,\ell}(x) \partial_y \textbf{J}_{r,\ell'}(y) & = \frac{2r}{(x - y)^3}\delta_{\ell+\ell ',0} + \cdots \, , \\
    \partial_x \textbf{J}_{r,\ell}(x) \partial_y \textbf{J}_{r,\ell'}(y) & =- \frac{6r }{(x - y)^4}\delta_{\ell+\ell ',0} + \cdots \, .
\end{align}
\end{subequations}

Now that we have this current operator, we come to defining the stress tensor $\textbf{T}$ corresponding to the generator~\eqref{eq:Lk-1_op_gen},
\begin{align}
    \textbf{T}_r(x) = 
    \sum_{\ell = 0}^{\lfloor r/2 \rfloor} \textbf{T}_{r,\ell}(x)
    \, ,
\end{align}
where each part is given by
\begin{align}
    \textbf{T}_{r,\ell}(x) =
    \begin{cases}
    \displaystyle
    \frac{1}{2r} {: \textbf{J}_{r,0} \textbf{J}_{r,0} :}(x) + \frac{Q}{\sqrt{2}r} \partial_x \textbf{J}_{r,0}(x)
    & (\ell = 0) \\[1em]
    \displaystyle
    \frac{1}{2r} {: \textbf{J}_{r,r/2} \textbf{J}_{r,r/2} :}(x)
    & (\ell = r/2, 2 \mathbb{N}) \\[1em] \displaystyle
    \frac{1}{r} {: \textbf{J}_{r,\ell} \textbf{J}_{r,-\ell } :}(x)
    & (\text{otherwise})
    \end{cases}
\end{align}
Applying the OPEs for the currents~\eqref{eq:JJ_OPE}, we obtain the OPE for the stress tensor,
\begin{align}
    \textbf{T}_{r,\ell}(x) \textbf{T}_{r,\ell}(y) = \frac{c_{r,\ell}/2}{(x - y)^4} + \frac{2 \textbf{T}_{r,\ell}(y)}{(x - y)^2} + \frac{\partial_y \textbf{T}_{r,\ell}(y)}{x - y} + \cdots
    \, ,
\end{align}
which reproduces the central charge $c_{r,\ell}$ shown in \eqref{eq:cc}.

\section{Discussion}
\label{sec:Discussion}

In this paper, we have discussed the root of unity limit of the $\textbf{q}$-Virasoro matrix model, and explored the associated Virasoro constraint.
We have clarified that the corresponding central charge agrees with the conjectural result obtained from the degree $r$ generalization of the minimal model CFT perspective.

We address several issues along the direction discussed in this paper.
The first is the origin of the Uglov matrix model itself. 
In this paper, we have formally defined the Uglov matrix model from the root of unity limit of the $\textbf{q}$-deformed matrix model.
Since the $\textbf{q}$-matrix model has the interpretation as the 3d gauge theory partition function, it would be possible that the Uglov matrix model also has a similar gauge theory interpretation.
From this point of view, recalling that the Uglov limit of the 5d gauge theory on $\mathbb{C}^2 \times \mathbb{S}^1$ is reduced to the 4d theory on the orbifold $\mathbb{C}^2/\mathbb{Z}_r$ (or its resolution $\widetilde{\mathbb{C}^2/\mathbb{Z}_r}$), it is expected that the Uglov matrix model may have a link with the 2d gauge theory on the orbifold, $\mathbb{C}/\mathbb{Z}_r$.
It would be interesting to derive the Uglov matrix model from the gauge theory perspective on the 2d orbifold as discussed in~\cite{Kimura:2011wh,Fujimori:2012ab}.

The second is a generalization of the Uglov-Virasoro constraint to the W-constraint.
As is known that the W-constraint is naturally obtained from the multi-matrix model, it seems to be reasonable to study the multi-matrix version of the Uglov matrix model.
In fact, the original conjecture of the central charge given in~\cite{Kimura:2012sc} is
\begin{align}
    c = r (N-1) - \frac{N(N^2-1)}{r} \mathcal{Q}_\beta^2
    \, ,
\end{align}
for the associated W$_N$ algebra (The Virasoro algebra corresponds to $N = 2$).
It would be interesting to reproduce this central charge from the W-constraint of the multi-matrix analog of the Uglov matrix model.

\subsubsection*{Acknowledgments}

This work was supported in part by ``Investissements d'Avenir'' program, Project ISITE-BFC (No.~ANR-15-IDEX-0003), EIPHI Graduate School (No.~ANR-17-EURE-0002), and Bourgogne-Franche-Comté region.

\bibliographystyle{amsalpha_mod}
\bibliography{ref}

\newcommand{\etalchar}[1]{$^{#1}$}
\providecommand{\bysame}{\leavevmode\hbox to3em{\hrulefill}\thinspace}
\providecommand{\MR}{\relax\ifhmode\unskip\space\fi MR }
\providecommand{\MRhref}[2]{%
  \href{http://www.ams.org/mathscinet-getitem?mr=#1}{#2}
}
\providecommand{\href}[2]{#2}
\begin{thebibliography}{AMOS95}

\bibitem[AGT10]{Alday:2009aq}
L.~F. Alday, D.~Gaiotto, and Y.~Tachikawa, \emph{{Liouville Correlation
  Functions from Four-dimensional Gauge Theories}},
  \href{https://doi.org/10.1007/s11005-010-0369-5}{Lett. Math. Phys.
  \textbf{91} (2010)}, 167--197,
  \href{https://arxiv.org/abs/0906.3219}{{\ttfamily arXiv:0906.3219 [hep-th]}}.

\bibitem[AMOS95]{Awata:1994xd}
H.~Awata, Y.~Matsuo, S.~Odake, and J.~Shiraishi, \emph{{Collective field
  theory, Calogero-Sutherland model and generalized matrix models}},
  \href{https://doi.org/10.1016/0370-2693(95)00055-P}{Phys. Lett. B
  \textbf{347} (1995)}, 49--55,
  \href{https://arxiv.org/abs/hep-th/9411053}{{\ttfamily
  arXiv:hep-th/9411053}}.

\bibitem[BDP14]{Beem:2012mb}
C.~Beem, T.~Dimofte, and S.~Pasquetti, \emph{{Holomorphic Blocks in Three
  Dimensions}}, \href{https://doi.org/10.1007/JHEP12(2014)177}{JHEP \textbf{12}
  (2014)}, 177, \href{https://arxiv.org/abs/1211.1986}{{\ttfamily
  arXiv:1211.1986 [hep-th]}}.

\bibitem[BF11]{Belavin:2011pp}
V.~Belavin and B.~Feigin, \emph{{Super Liouville conformal blocks from
  $\mathcal{N}=2$ SU(2) quiver gauge theories}},
  \href{https://doi.org/10.1007/JHEP07(2011)079}{JHEP \textbf{07} (2011)}, 079,
  \href{https://arxiv.org/abs/1105.5800}{{\ttfamily arXiv:1105.5800 [hep-th]}}.

\bibitem[BP98]{Bouwknegt:1998da}
P.~Bouwknegt and K.~Pilch, \emph{{On deformed W algebras and quantum affine
  algebras}}, \href{https://doi.org/10.4310/ATMP.1998.v2.n2.a6}{Adv. Theor.
  Math. Phys. \textbf{2} (1998)}, 357--397,
  \href{https://arxiv.org/abs/math/9801112}{{\ttfamily arXiv:math/9801112}}.

\bibitem[CLPZ19]{Cassia:2019sjk}
L.~Cassia, R.~Lodin, A.~Popolitov, and M.~Zabzine, \emph{{Exact SUSY Wilson
  loops on S$^{3}$ from $q$-Virasoro constraints}},
  \href{https://doi.org/10.1007/JHEP12(2019)121}{JHEP \textbf{12} (2019)}, 121,
  \href{https://arxiv.org/abs/1909.10352}{{\ttfamily arXiv:1909.10352
  [hep-th]}}.

\bibitem[CLZ20]{Cassia:2020uxy}
L.~Cassia, R.~Lodin, and M.~Zabzine, \emph{{On matrix models and their
  $q$-deformations}}, \href{https://doi.org/10.1007/JHEP10(2020)126}{JHEP
  \textbf{10} (2020)}, 126, \href{https://arxiv.org/abs/2007.10354}{{\ttfamily
  arXiv:2007.10354 [hep-th]}}.

\bibitem[CZ21]{Cassia:2021uly}
L.~Cassia and M.~Zabzine, \emph{{On refined Chern-Simons and refined ABJ matrix
  models}}, \href{https://arxiv.org/abs/2107.07525}{{\ttfamily arXiv:2107.07525
  [hep-th]}}.

\bibitem[DVV91]{Dijkgraaf:1990rs}
R.~Dijkgraaf, H.~L. Verlinde, and E.~P. Verlinde, \emph{{Loop equations and
  Virasoro constraints in nonperturbative 2-D quantum gravity}},
  \href{https://doi.org/10.1016/0550-3213(91)90199-8}{Nucl. Phys. B
  \textbf{348} (1991)}, 435--456.

\bibitem[EB12]{Estienne:2012si}
B.~Estienne and B.~A. Bernevig, \emph{{Spin-singlet quantum Hall states and
  Jack polynomials with a prescribed symmetry}},
  \href{https://doi.org/10.1016/j.nuclphysb.2011.12.007}{Nucl. Phys. B
  \textbf{857} (2012)}, 185--206,
  \href{https://arxiv.org/abs/1107.2534}{{\ttfamily arXiv:1107.2534
  [cond-mat.str-el]}}.

\bibitem[FKNO12]{Fujimori:2012ab}
T.~Fujimori, T.~Kimura, M.~Nitta, and K.~Ohashi, \emph{{Vortex counting from
  field theory}}, \href{https://doi.org/10.1007/JHEP06(2012)028}{JHEP
  \textbf{06} (2012)}, 028, \href{https://arxiv.org/abs/1204.1968}{{\ttfamily
  arXiv:1204.1968 [hep-th]}}.

\bibitem[FL88]{Fateev:1987zh}
V.~A. Fateev and S.~L. Lukyanov, \emph{{The Models of Two-Dimensional Conformal
  Quantum Field Theory with Z$_N$ Symmetry}},
  \href{https://doi.org/10.1142/S0217751X88000205}{Int. J. Mod. Phys. A
  \textbf{3} (1988)}, 507.

\bibitem[FZ85]{Fateev:1985mm}
V.~A. Fateev and A.~B. Zamolodchikov, \emph{{Nonlocal (parafermion) currents in
  two-dimensional conformal quantum field theory and self-dual critical points
  in Z$_N$-symmetric statistical systems}},
  \href{http://83.149.229.155/cgi-bin/e/index/e/62/2/p215?a=list}{Sov. Phys.
  JETP \textbf{62} (1985)}, 215--225.

\bibitem[IOY13]{Itoyama:2013mca}
H.~Itoyama, T.~Oota, and R.~Yoshioka, \emph{{2d-4d Connection between
  $q$-Virasoro/W Block at Root of Unity Limit and Instanton Partition Function
  on ALE Space}}, \href{https://doi.org/10.1016/j.nuclphysb.2013.10.012}{Nucl.
  Phys. B \textbf{877} (2013)}, 506--537,
  \href{https://arxiv.org/abs/1308.2068}{{\ttfamily arXiv:1308.2068 [hep-th]}}.

\bibitem[IOY14]{Itoyama:2014pca}
\bysame, \emph{{$q$-Virasoro/W Algebra at Root of Unity and Parafermions}},
  \href{https://doi.org/10.1016/j.nuclphysb.2014.10.006}{Nucl. Phys. B
  \textbf{889} (2014)}, 25--35,
  \href{https://arxiv.org/abs/1408.4216}{{\ttfamily arXiv:1408.4216 [hep-th]}}.

\bibitem[Kim11]{Kimura:2011zf}
T.~Kimura, \emph{{Matrix model from $\mathcal{N} = 2$ orbifold partition
  function}}, \href{https://doi.org/10.1007/JHEP09(2011)015}{JHEP \textbf{09}
  (2011)}, 015, \href{https://arxiv.org/abs/1105.6091}{{\ttfamily
  arXiv:1105.6091 [hep-th]}}.

\bibitem[Kim12a]{Kimura:2011gq}
\bysame, \emph{{$\beta$-ensembles for toric orbifold partition function}},
  \href{https://doi.org/10.1143/PTP.127.271}{Prog. Theor. Phys. \textbf{127}
  (2012)}, 271--285, \href{https://arxiv.org/abs/1109.0004}{{\ttfamily
  arXiv:1109.0004 [hep-th]}}.

\bibitem[Kim12b]{Kimura:2012sc}
\bysame, \emph{{Spinless basis for spin-singlet FQH states}},
  \href{https://doi.org/10.1143/PTP.128.829}{Prog. Theor. Phys. \textbf{128}
  (2012)}, 829--843, \href{https://arxiv.org/abs/1201.1903}{{\ttfamily
  arXiv:1201.1903 [cond-mat.mes-hall]}}.

\bibitem[KK09]{Kuramoto:2009}
Y.~Kuramoto and Y.~Kato,
  \href{https://doi.org/10.1017/CBO9780511596827}{\emph{{Dynamics of
  One-Dimensional Quantum Systems: Inverse-Square Interaction Models}}},
  Cambridge University Press, 2009.

\bibitem[KN11]{Kimura:2011wh}
T.~Kimura and M.~Nitta, \emph{{Vortices on Orbifolds}},
  \href{https://doi.org/10.1007/JHEP09(2011)118}{JHEP \textbf{09} (2011)}, 118,
  \href{https://arxiv.org/abs/1108.3563}{{\ttfamily arXiv:1108.3563 [hep-th]}}.

\bibitem[KN21]{Kimura:2021ngu}
T.~Kimura and F.~Nieri, \emph{{Intersecting Defects and Supergroup Gauge
  Theory}}, \href{https://doi.org/10.1088/1751-8121/ac2716}{J. Phys. A: Math.
  Theor. \textbf{54} (2021)}, 435401,
  \href{https://arxiv.org/abs/2105.02776}{{\ttfamily arXiv:2105.02776
  [hep-th]}}.

\bibitem[KP19]{Kimura:2019xzj}
T.~Kimura and V.~Pestun, \emph{{Twisted reduction of quiver W-algebras}},
  \href{https://arxiv.org/abs/1905.03865}{{\ttfamily arXiv:1905.03865
  [hep-th]}}.

\bibitem[LPSZ20]{Lodin:2018lbz}
R.~Lodin, A.~Popolitov, S.~Shakirov, and M.~Zabzine, \emph{{Solving
  $q$-Virasoro constraints}},
  \href{https://doi.org/10.1007/s11005-019-01216-5}{Lett. Math. Phys.
  \textbf{110} (2020)}, no.~1, 179--210,
  \href{https://arxiv.org/abs/1810.00761}{{\ttfamily arXiv:1810.00761
  [hep-th]}}.

\bibitem[MM90]{Mironov:1990im}
A.~Mironov and A.~Morozov, \emph{{On the origin of Virasoro constraints in
  matrix models: Lagrangian approach}},
  \href{https://doi.org/10.1016/0370-2693(90)91078-P}{Phys. Lett. B
  \textbf{252} (1990)}, 47--52.

\bibitem[Mor94]{Morozov:1994hh}
A.~Morozov, \emph{{Integrability and matrix models}},
  \href{https://doi.org/10.1070/PU1994v037n01ABEH000001}{Phys. Usp. \textbf{37}
  (1994)}, 1--55, \href{https://arxiv.org/abs/hep-th/9303139}{{\ttfamily
  arXiv:hep-th/9303139}}.

\bibitem[Nek16]{Nekrasov:2015wsu}
N.~Nekrasov, \emph{{BPS/CFT correspondence: non-perturbative Dyson-Schwinger
  equations and qq-characters}},
  \href{https://doi.org/10.1007/JHEP03(2016)181}{JHEP \textbf{03} (2016)}, 181,
  \href{https://arxiv.org/abs/1512.05388}{{\ttfamily arXiv:1512.05388
  [hep-th]}}.

\bibitem[NNZ17]{Nedelin:2016gwu}
A.~Nedelin, F.~Nieri, and M.~Zabzine, \emph{{$q$-Virasoro modular double and 3d
  partition functions}},
  \href{https://doi.org/10.1007/s00220-017-2882-1}{Commun. Math. Phys.
  \textbf{353} (2017)}, no.~3, 1059--1102,
  \href{https://arxiv.org/abs/1605.07029}{{\ttfamily arXiv:1605.07029
  [hep-th]}}.

\bibitem[NPP15]{Nieri:2013yra}
F.~Nieri, S.~Pasquetti, and F.~Passerini, \emph{{3d and 5d Gauge Theory
  Partition Functions as $q$-deformed CFT Correlators}},
  \href{https://doi.org/10.1007/s11005-014-0727-9}{Lett. Math. Phys.
  \textbf{105} (2015)}, no.~1, 109--148,
  \href{https://arxiv.org/abs/1303.2626}{{\ttfamily arXiv:1303.2626 [hep-th]}}.

\bibitem[NT11]{Nishioka:2011jk}
T.~Nishioka and Y.~Tachikawa, \emph{{Central charges of para-Liouville and Toda
  theories from M-5-branes}},
  \href{https://doi.org/10.1103/PhysRevD.84.046009}{Phys. Rev. D \textbf{84}
  (2011)}, 046009, \href{https://arxiv.org/abs/1106.1172}{{\ttfamily
  arXiv:1106.1172 [hep-th]}}.

\bibitem[NZ17]{Nedelin:2015mio}
A.~Nedelin and M.~Zabzine, \emph{{$q$-Virasoro constraints in matrix models}},
  \href{https://doi.org/10.1007/JHEP03(2017)098}{JHEP \textbf{03} (2017)}, 098,
  \href{https://arxiv.org/abs/1511.03471}{{\ttfamily arXiv:1511.03471
  [hep-th]}}.

\bibitem[Pas12]{Pasquetti:2011fj}
S.~Pasquetti, \emph{{Factorisation of $\mathcal{N} = 2$ Theories on the
  Squashed 3-Sphere}}, \href{https://doi.org/10.1007/JHEP04(2012)120}{JHEP
  \textbf{04} (2012)}, 120, \href{https://arxiv.org/abs/1111.6905}{{\ttfamily
  arXiv:1111.6905 [hep-th]}}.

\bibitem[Pas17]{Pasquetti:2016dyl}
\bysame, \emph{{Holomorphic blocks and the 5d AGT correspondence}},
  \href{https://doi.org/10.1088/1751-8121/aa60fe}{J. Phys. A \textbf{50}
  (2017)}, no.~44, 443016, \href{https://arxiv.org/abs/1608.02968}{{\ttfamily
  arXiv:1608.02968 [hep-th]}}.

\bibitem[Pes12]{Pestun:2007rz}
V.~Pestun, \emph{{Localization of gauge theory on a four-sphere and
  supersymmetric Wilson loops}},
  \href{https://doi.org/10.1007/s00220-012-1485-0}{Commun. Math. Phys.
  \textbf{313} (2012)}, 71--129,
  \href{https://arxiv.org/abs/0712.2824}{{\ttfamily arXiv:0712.2824 [hep-th]}}.

\bibitem[PZ{\etalchar{+}}17]{Pestun:2016zxk}
V.~Pestun, M.~Zabzine, et~al., \emph{{Localization techniques in quantum field
  theories}}, \href{https://doi.org/10.1088/1751-8121/aa63c1}{J. Phys. A
  \textbf{50} (2017)}, no.~44, 440301,
  \href{https://arxiv.org/abs/1608.02952}{{\ttfamily arXiv:1608.02952
  [hep-th]}}.

\bibitem[SKAO96]{Shiraishi:1995rp}
J.~Shiraishi, H.~Kubo, H.~Awata, and S.~Odake, \emph{{A Quantum deformation of
  the Virasoro algebra and the Macdonald symmetric functions}},
  \href{https://doi.org/10.1007/BF00398297}{Lett. Math. Phys. \textbf{38}
  (1996)}, 33--51, \href{https://arxiv.org/abs/q-alg/9507034}{{\ttfamily
  arXiv:q-alg/9507034}}.

\bibitem[Ugl98]{Uglov:1997ia}
D.~Uglov, \emph{{Yangian Gelfand-Zetlin bases, $\mathfrak{gl}_N$ Jack
  polynomials and computation of dynamical correlation functions in the spin
  Calogero-Sutherland model}},
  \href{https://doi.org/10.1007/s002200050283}{Commun. Math. Phys. \textbf{193}
  (1998)}, 663--696, \href{https://arxiv.org/abs/hep-th/9702020}{{\ttfamily
  arXiv:hep-th/9702020}}.

\bibitem[Wyl09]{Wyllard:2009hg}
N.~Wyllard, \emph{{$A_{N-1}$ conformal Toda field theory correlation functions
  from conformal $\mathcal{N} = 2$ SU($N$) quiver gauge theories}},
  \href{https://doi.org/10.1088/1126-6708/2009/11/002}{JHEP \textbf{11}
  (2009)}, 002, \href{https://arxiv.org/abs/0907.2189}{{\ttfamily
  arXiv:0907.2189 [hep-th]}}.

\bibitem[YS20]{Yoshida:2014ssa}
Y.~Yoshida and K.~Sugiyama, \emph{{Localization of three-dimensional
  $\mathcal{N}=2$ supersymmetric theories on $S^1 \times D^2$}},
  \href{https://doi.org/10.1093/ptep/ptaa136}{PTEP \textbf{2020} (2020)},
  no.~11, 113B02, \href{https://arxiv.org/abs/1409.6713}{{\ttfamily
  arXiv:1409.6713 [hep-th]}}.

\end{thebibliography}

\end{document}